\documentstyle[12pt]{article}

\def\baselinestretch{1.5}
\def\thefootnote{\fnsymbol{footnote}}
\input{epsf}
\newcommand{\mysection}{\setcounter{equation}{0}\section}
\renewcommand{\thefootnote}{\fnsymbol{footnote}}
\renewcommand{\theequation}{\thesection.\arabic{equation}}

\newcommand{\startappendix}{
\renewcommand{\thesection}{\Alph{section}}}

\begin{document}
\begin{titlepage}
\begin{center}

\today \hfill arch-ive/9411132\\

\vskip .2in

{\Large \bf  Correlation Length and Average Loop Length
                   of the Fully-Packed Loop Model}

\vskip .5in

Anton Kast\footnote{\tt anton@physics.berkeley.edu}\\

{\small {\em
Department of Physics \\
University of California at Berkeley \\
Berkeley, California 94720, USA}}

\end{center}
\vskip .5in

\begin{abstract}

The fully-packed loop model of closed paths covering the honeycomb lattice
is studied through its identification with the $sl_q(3)$ integrable lattice
model.  Some known results from the Bethe ansatz solution of this model are
reviewed.  The free energy, correlation length, and the ensemble average loop
length are given explicitly for the many-loop phase.  The results are compared
with the known result for the model's surface tension.  A perturbative
formalism is introduced and used to verify results.

\end{abstract}
\end{titlepage}
\newpage
\renewcommand{\thepage}{\arabic{page}}
\setcounter{page}{1}

\section {Review of the FPL model}
\setcounter{equation}{0}
\baselineskip 18.5pt

\begin{figure}
\epsfbox{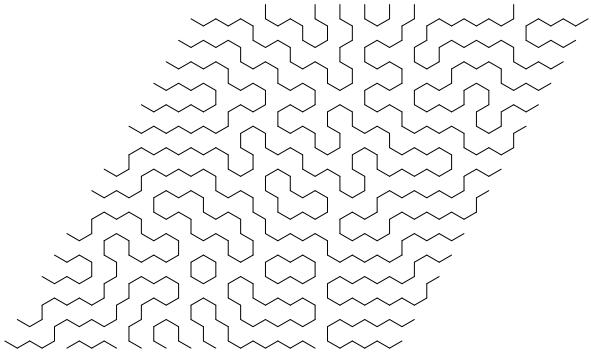}
\caption{An example of a fully-packed loop configuration.  With periodic
boundary conditions, this configuration has 11 loops.}
\label{example}
\end{figure}

In a recent article~\cite{Blote} Bl\"ote and Nienhuis performed numerical
investigations of what they termed the fully-packed loop (FPL) model.  This is
a statistical model where the ensemble is the set of all combinations of closed
paths on the honeycomb lattice that visit every vertex and do not intersect.
The Boltzmann weight of such a filling set of paths is just the exponential of
the number of paths, i.e. the energy of a configuration is the number of closed
loops used to cover the lattice.  An example of a  fully-packed configuration
of loops on this lattice is shown in Figure~\ref{example}.  The partition
function for this model may be represented as
\begin{equation}
Z_{FPL}(n)=\sum_{C}n^{P(C)}
\label{ZFPL}
\end{equation}
where the sum is over all $C$, the coverings of the vertices of the hexagonal
lattice by closed nonintersecting paths,  $P(C)$ is the number of paths in the
covering $C$, and $n$ is a generalized activity. This model was originally
studied for its interest as the low-temperature limit of the $O(n)$ vector
lattice models~\cite{Nienhuis,Domany}.  In this limit, the dimensionality of
vectors $n$ is just the activity $n$ in equation~(\ref{ZFPL}).

The partition function~(\ref{ZFPL}) is apparently the generating function for
the numbers of ways to cover the hexagonal lattice by any number of closed
paths. Its calculation in the thermodynamic limit is an interesting
combinatorial problem.

More recently, Batchelor, Suzuki and Yung~\cite{Batchelor} pointed out that
previous authors~\cite{Warnaar,Reshetikhin} had exploited an identification of
the FPL model with the integrable lattice model associated to the quantum group
$sl_q(3)$~\cite{BdVV}.  This integrable model is a vertex model on the
square lattice where each link of the lattice can be in one of three states,
and the vertex weights are given by the R-matrix for $sl_q(3)$ as in
Figure~\ref{weights}.  The R-matrix depends on a deformation parameter $\gamma
= \log(q)$, as well as a spectral parameter $\theta$ typical of integrable
theories.  Denoting the $sl_q(3)$ partition function as $Z_{sl_q(3)}(\gamma,
\theta)$, the precise idenfication is
\begin{equation}
Z_{FPL}(e^\gamma + e^{-\gamma}) = \left.
(e^\gamma-e^{-\gamma})^{-N} Z_{sl_q(3)}(\gamma,\theta) \right|_{\theta=-\gamma}
\label{identification}
\end{equation}
where $N$ is the volume of the lattice (the number of hexagonal faces).  Since
the model is integrable, much exact information can be derived.  In particular,
the model's Bethe equations have been constructed and solved.

\begin{figure}
\begin{center}
\begin{picture}(400,100)
\put(50,25){\line(0,1){50}}
\put(25,50){\line(1,0){50}}
\put(50,25){a} \put(50,75){b} \put(25,50){i} \put(75,50){j}

\put(100,50) { $ \begin{array}{ll}
                 = & R^{ij}_{ab}(\gamma,\theta)  \\
                 = & \delta^i_b \, \delta^j_a \,
                     \frac{\sinh(\theta)}{\sinh(\theta+\gamma)} \; + \\
                   & \delta^i_a \, \delta^j_b
                     \left[ (1-\delta_{ab})
                     \frac{\sinh(\gamma)}{\sinh(\theta+\gamma)}
                     e^{\Delta_{ab}\theta}
                     + \delta_{ab}
                       \left( 1-\frac{\sinh(\theta)}{\sinh(\theta+\gamma)}
                       \right)
                     \right] \\
\end{array} $ }

\end{picture}
\end{center}
\caption{The definition of vertex weights in the integrable lattice model
associated to $sl_q(d+1)$.  $\gamma$ and $\theta$ are parameters of the model.
States on edges are labelled by roman indices with $(d+1)$ possible values.  In
the formula, $\Delta_{ab} = \frac{a-b}{|a-b|} + 2 \frac{b-a}{d+1}$.}
\label{weights}
\end{figure}

One of the more important results that have been derived in this way is the
existence of a phase transition in the model~(\ref{ZFPL}) at
$n=2$~\cite{Reshetikhin,Baxter1970}.  At larger $n$, larger numbers of loops
are favored and at smaller $n$ configurations with fewer loops are favored.  It
has been conjectured that this transition is between a large-$n$ phase where
the average loop length is finite and a small-$n$ phase where this average is
infinite.

In section~\ref{FLsection} a simple relation between the free energy of the FPL
model and the ensemble average length of loops is derived.  From the known
solution to the Bethe equations of the $sl_q(3)$ integrable lattice model, the
free energy is identified and used to graph the exact value of the average
loop length as a function of $n$.

Identifying $n=e^\gamma + e^{-\gamma}$, it becomes apparent that $n>2$
corresponds to the integrable model for $\gamma$ real and $0<n<2$ corresponds
to $\gamma$ purely imaginary.  The former phase is known~\cite{BdVV} to be
massive, in the sense that there is a gap in the spectrum of eigenvalues of the
transfer matrix between the leading eigenvalue and the next-leading
eigenvalue.  By standard arguments~\cite{BaxterBook}, this implies a finite
correlation length.  The gap tends to zero as $\gamma$ goes to zero, showing
that $n=2$ is a critical point of the model~(\ref{ZFPL}).

In section~\ref{xiSection} this correlation length is studied by considering
the spectrum of eigenvalues of this transfer matrix.  The spectrum may be
deduced directly from the model's Bethe equations.  In the case that the
transfer matrix is symmetric and therefore has real eigenvalues, the
correlation length is related to the maximum eigenvalue $\Lambda_{\rm max}$
and next-leading eigenvalue $\Lambda_1$ by
\begin{equation}
\xi^{-1} = \log \frac{\Lambda_{\rm max}}{\Lambda_1} .
\label{xiDef}
\end{equation}

\section {Review of $sl_q$ integrable models}
\setcounter{equation}{0}
\baselineskip 18.5pt

The R-matrix of the $sl_q(d+1)$ quantum group is an $(n+1)^2 \times (n+1)^2$
matrix that may be interpreted as a matrix of Boltzmann weights of a vertex
model on the square lattice as shown in Figure~\ref{weights}.  Since this
matrix satisfies the Yang-Baxter equation, the associated transfer matrix
commutes with itself evaluated at differing values of the spectral parameter
and the model is exactly solvable by a recursive set of $d$ nested Bethe
ans\"atze~\cite{BdVV}.

The formula for eigenvalues of the transfer matrix for the range of parameters,
$\theta > -\frac{1}{2}\gamma$ and $\gamma$ real and positive, is
\begin{equation}
\Lambda = \prod_k
           \frac{\sinh(i\lambda_k+\frac{1}{2}\gamma-\theta)}
                {\sinh(i\lambda_k-\frac{1}{2}\gamma-\theta)}
\label{eigenvalue}
\end{equation}
where the product is over roots, $\lambda_k$ of a set of Bethe equations and we
have neglected terms that do not contribute in the thermodynamic limit. In
this limit the $\lambda_k$ are distributed in the interval
$[-\pi/2,\pi/2]$ with the density,
\begin{equation}
\rho_{\rm max}(\lambda) = \sum_{m=-\infty}^{\infty} e^{2im\lambda}
   \frac{1}{\pi} \frac{\sinh(dm\gamma)}{\sinh[(d+1)m\gamma]}.
\end{equation}
For eigenvalues near the maximum eigenvalue, the changes in the distribution
are parameterized by the locations of holes $\theta^q_h$, $q=1 \cdots d$, $h=1
\cdots N_q$, according to
\begin{equation}
\rho(\lambda)-\rho_{\rm max}(\lambda) = \sum_{m=\infty}^{\infty} e^{2im\lambda}
   \sum_{q=1}^{d} \frac{-1}{\pi} e^{|m\gamma|}
   \frac{\sinh[(d+1-q)m\gamma]}{\sinh[(d+1)m\gamma]}
   \sum_{h=1}^{N_q} e^{-2im\theta^q_h}.
\label{density}
\end{equation}

The numbers of holes are constrained to satisfy the relation,
\begin{equation}
\sum_{q=1}^{d} q N_q = {\rm multiple\;of\;} (n+1).
\end{equation}

The formulas~(\ref{eigenvalue}),~(\ref{density}) may be combined in the
thermodynamic limit to yield a formula for eigenvalues of the transfer matrix
of the integrable model,
\begin{equation}
\log \Lambda = \int^{\pi/2}_{-\pi/2} \rho(\lambda)
  \log \left[ \frac{\sinh(i\lambda+\frac{1}{2}\gamma-\theta)}
                   {\sinh(i\lambda-\frac{1}{2}\gamma-\theta)} \right]
   \,d\lambda.
\label{formula}
\end{equation}

\section{Free energy and average loop length \label{FLsection}}
\setcounter{equation}{0}
\baselineskip 18.5pt

The formulas of the preceding section in the case $d=2$ yield directly the free
energy density of the model~(\ref{ZFPL}) for the $n>2$ phase as the logarithm
of the maximum eigenvalue of the transfer matrix, rescaled by the factor of
equation~(\ref{identification}). This free energy was actually derived in 1970
by Baxter~\cite{Baxter1970} as the solution to a weighted three-coloring
problem on the honeycomb lattice.  The free energy density of the FPL model in
the $n>2$ phase is
\begin{eqnarray}
F_{FPL}(n) & \equiv & \lim_{N \rightarrow \infty}
                      \frac{1}{N} \log Z_{FPL}(n) \nonumber \\
           & = & \log \left\{ q^{1/3} \prod_{m=1}^\infty
                 \frac{(1-q^{-6m+2})^2}{(1-q^{-6p+4})(1-q^{-6p})} \right\}
\label{FFPL}
\end{eqnarray}
where $n=q+q^{-1}$, and $q=e^\gamma>1$.  This function has an essential
singularity at $q=1$.  The free energy for $n<2$ and with periodic boundary
conditions is given in integral form in~\cite{Batchelor}.

It is interesting to note that for both phases, the free energy density gives
the ensemble average length of loops.  Since a configuration $C$ on a lattice
of
$N$ faces has $2N$ occupied links, the total length of loops is always $2N$.
The average loop length of configuration $C$ is therefore $2N/P(C)$.
If we define the ensemble average loop length $L_N(n)$  by
\begin{equation}
L_N(n) = \frac{1}{Z_{FPL}(n)} \sum_C \frac{2N}{P(C)} n^{P(C)}, \label{L}
\label{Ldef}
\end{equation}
then from inspection of equation~(\ref{ZFPL}) it is clear that
\begin{equation}
\frac{d}{dn} \left[ L_N(n) Z_{FPL}(n) \right] = \frac{2N}{n} Z_{FPL}(n).
\label{Leqn}
\end{equation}
The general solution to this equation can be written up to quadrature by direct
integration:
\begin{equation}
L_N(n) = \frac{1}{Z_{FPL}(n)} \int^n_C \frac{2N}{n'} Z_{FPL}(n') dn'
\end{equation}
where the lower limit of integration is an undetermined constant. In terms of
the free energy density $F_{FPL} = (1/N) \log Z_{FPL}$, this becomes
\begin{equation}
L_N(n) = 2N e^{-N F_{FPL}(n)} \int^n_C \frac{e^{N F_{FPL}(n')}}{n'} dn'.
\label{Lintegral}
\end{equation}
The integral in equation~(\ref{Lintegral}) can be evaluated by steepest
descent.  The result is
\begin{equation}
L_N(n) = \frac{2N}{n}
         \exp \left( -Nn \frac{dF_{FPL}}{dn}(n) \right)
         \left[ \frac{\exp \left[ Nn'\frac{dF_{FPL}}{dn}(n') \right] }
                {N \frac{dF_{FPL}}{dn}(n')} \right]^n_C.
\end{equation}
The constant of integration may now be determined from the known value of
$L_N(n)$ at $n \rightarrow \infty$.  As will be shown in
section~\ref{perturbSection}, in this limit $Z_N(n) \simeq 3 n^{N/3}$,
$\frac{dF_{FPL}}{dn}(n) \simeq 1/3n$, and $L_N(n) = 6$.  These imply that  $C =
- \infty$, so in the thermodynamic limit
\begin{equation}
L_N(n) = \frac{2}{n \frac{dF_{FPL}}{dn}}.
\end{equation}
In this calculation we have neglected corrections of order $1/N$ to $L_N(n)$.

\begin{figure}
\epsfbox{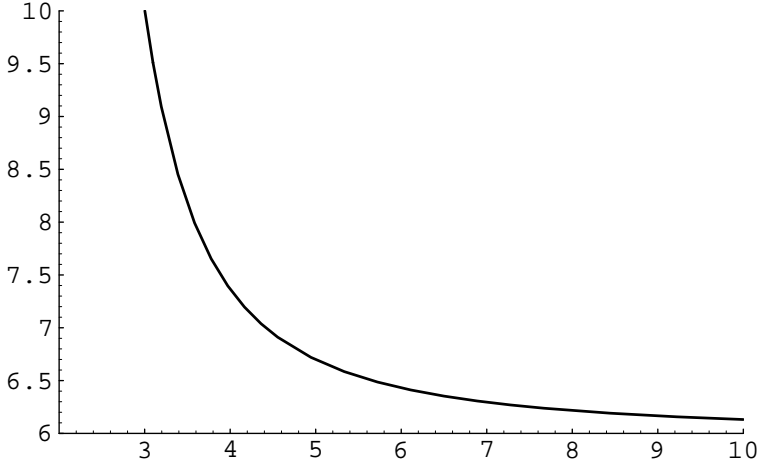}
\caption{This is a graph of the average loop length of the FPL model versus
$n$, the fugacity of loops.  The critical point is at $n=2$.}
\label{lengthGraph}
\end{figure}

A graph of the ensemble average loop length versus $n$ in the large-$n$ phase
is shown in Figure~\ref{lengthGraph}.  This verifies the conjecture
of~\cite{Reshetikhin} that loop length diverges at the critical point.

\section {Correlation length \label{xiSection}}
\setcounter{equation}{0}
\baselineskip 18.5pt

To obtain the correlation length, we must compute the
expression~(\ref{formula}) for the minimal hole distribution.  When $d=2$,
there are two choices for the $N_q$.  Either $N_1=3$ and $N_2=0$, or $N_1=1$
and $N_2=1$.  In each case, the eigenvalue gap is minimized for holes at
$\theta^q_h=\frac{\pi}{2}$ where the sum in equation~(\ref{density}) after
integration in~(\ref{formula}) is oscillatory.  The transfer matrix of the
model is symmetric at the point $\theta=-\frac{1}{2}\gamma$, and its
eigenvalues are then real.  After setting $\theta$ to this value, the
correlation length of the model is given by equation~(\ref{xiDef}).

Considering the case $N_1=3$ and $N_2=0$, we denote the next-leading eigenvalue
for this hole distribution as $\Lambda_{30}$.  The equation~(\ref{formula})
together with the formula for densities~(\ref{density}) gives
\begin{equation}
\log \frac{\Lambda_{30}}{\Lambda_{\rm max}} =
   \sum_{m=-\infty}^{\infty} \phi_m \left(-\frac{1}{\pi}\right) e^{|m\gamma|}
   3 (-1)^m \frac{\sinh(2m\gamma)}{\sinh(3m\gamma)}
\label{formula2}
\end{equation}
where $\phi_m$ are the integrals over roots,
\begin{equation}
\phi_m \equiv \int^{\pi/2}_{-\pi/2} e^{2im\lambda}
  \log \left[ \frac{\sinh(i\lambda+\frac{1}{2}\gamma-\theta)}
                   {\sinh(i\lambda-\frac{1}{2}\gamma-\theta)} \right]
  \,d\lambda.
\label{integral}
\end{equation}

The integral in equation~(\ref{integral}) can easily be  performed by contour
integration.  After introducing the variables $q=e^\gamma$ and $z=e^\theta$,
the result for $-\frac{1}{2}\gamma < \theta < 0$ is
\begin{equation}
\int^{\pi/2}_{-\pi/2} e^{2im\lambda}
  \log \left[ \frac{\sinh(i\lambda+\frac{1}{2}\gamma-\theta)}
                   {\sinh(i\lambda-\frac{1}{2}\gamma-\theta)} \right]
  \,d\lambda =
\left\{ \begin{array}{ll}
        \frac{\pi}{m} [ 1-(z^2 q^{-1})^m ], & m>0 \\
        -2\pi\log z, & m=0 \\
        \frac{\pi}{m} [ 1-(z^2 q)^m ], & m<0
        \end{array} \right. .
\end{equation}
Substituting this result into equation~(\ref{formula2}) gives
\begin{equation}
\log \frac{\Lambda_{30}}{\Lambda_{\rm max}} =
   2 \log z + 3 \sum_{m>0} \frac{(-1)^m}{m} (z^{2m} - z^{-2m})
   \left( \frac{q^{2m}-q^{-2m}}{q^{3m}-q^{-3m}} \right) .
\end{equation}
After expanding the demoninator of the summand in a power series in $q^{-1}$,
this may be resummed to the form,
\begin{equation}
\log \frac{\Lambda_{30}}{\Lambda_{\rm max}} =
   2 \log z - 3 \sum_{m \geq 0}
   \log \left[ \frac{(1+z^2q^{-1}q^{-6m})(1+z^{-2}q^{-5}q^{-6m})}
                    {(1+z^{-2}q^{-1}q^{-6m})(1+z^2q^{-5}q^{-6m})} \right].
\end{equation}
This form is now convergent at the symmetric point,
$\theta = - \frac{1}{2}\gamma$ or equivalently $z^2=q^{-1}$.  We may therefore
evaluate it there to obtain the correlation length according to
equation~(\ref{xiDef}),
\begin{equation}
\xi^{-1} = 3 \log \left\{
   q^{1/3} \prod_{m>0} \frac{(1+q^2q^{-6m})(1+q^4q^{-6m})}
                            {(1+q^{-6m})(1+q^6q^{-6m})}
\right\}
\label{xi}
\end{equation}
This is the desired result, the correlation length of the FPL model where
$n=q+q^{-1}$ and $q>1$, or equivalently $q=+\sqrt{n^2-4}$.

The other possible choice of holes, $N_1=1$ and $N_2=1$ may be computed in
the same way to give
\begin{equation}
\log \frac{\Lambda_{\rm max}}{\Lambda_{11}} =
   \log \left\{ q \prod_{m>0}
   \frac{(1+q^2q^{-6m})(1+q^3q^{-6m})(1+q^3q^{-6m})(1+q^4q^{-6m})}
        {(1+q^{-6m})(1+qq^{-6m})(1+q^5q^{-6m})(1+q^6q^{-6m})} \right\} .
\label{other}
\end{equation}
This quantity is greater than~(\ref{xi}) for all $q>1$, so it is not the
inverse correlation length. For large $q$, the inequality may be seen by
considering the limiting forms of expressions~(\ref{xi}) and~(\ref{other}).
Rigorously, the multiplicands in~(\ref{other}) may be seen to be greater than
those in ~(\ref{xi}) term by term in $m$.

\section {Perturbative analysis \label{perturbSection}}
\setcounter{equation}{0}
\baselineskip 18.5pt

The FPL model has a natural large-$n$ expansion which allows simple
perturbative verifications of results.

When $n$ is large, the dominant configurations are those with large numbers of
loops.  The perturbative procedure is to approximate the sum over states by
including the configurations with the highest numbers of loops.

On a hexagonal lattice with number of faces $N$ a multiple of three, there
are three configurations with the maximum possible number of loops.  In these
states, one out of every three faces has a small loop around it and these
small loops lie on a triangular lattice.  A sample is shown in
Figure~\ref{noDefects}.  These three configurations differ by translations
and each has $N/3$ loops.

\newcommand{\edgeA}{\put(20,0){\line(-3,5){10}}}
\newcommand{\edgeB}{\put(20,0){\line(-3,-5){10}}}
\newcommand{\edgeC}{\put(10,-17){\line(-1,0){20}}}
\newcommand{\edgeD}{\put(-20,0){\line(3,-5){10}}}
\newcommand{\edgeE}{\put(-20,0){\line(3,5){10}}}
\newcommand{\edgeF}{\put(-10,17){\line(1,0){20}}}

\newcommand{\hexagon}{\edgeA \edgeB \edgeC \edgeD \edgeE \edgeF}

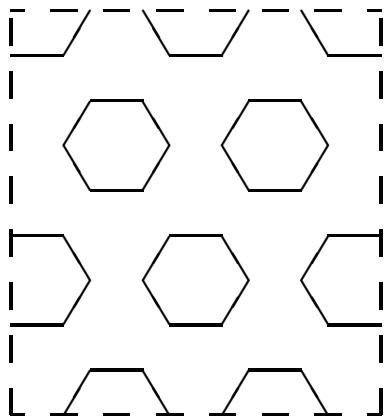
\begin{figure}
\begin{center}
\begin{picture}(140,153)

\thicklines

\put(20,0){\dashbox{10}(140,153){}}

\put(60,0){\edgeE \edgeF \edgeA}
\put(120,0){\edgeE \edgeF \edgeA}
\put(30,51){\edgeF \edgeA \edgeB \edgeC}
\put(90,51){\hexagon}
\put(150,51){\edgeC \edgeD \edgeE \edgeF}
\put(60,102){\hexagon}
\put(120,102){\hexagon}
\put(30,153){\edgeB \edgeC}
\put(90,153){\edgeB \edgeC \edgeD}
\put(150,153){\edgeC \edgeD}

\end{picture}
\caption{A sample from a configuration with the maximum number of loops.}
\label{noDefects}
\end{center}
\end{figure}

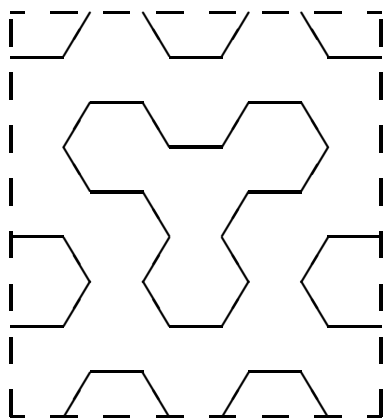
\begin{figure}
\begin{center}
\begin{picture}(140,153)

\thicklines

\put(20,0){\dashbox{10}(140,153){}}
\put(60,0){\edgeE \edgeF \edgeA}
\put(120,0){\edgeE \edgeF \edgeA}
\put(30,51){\edgeF \edgeA \edgeB \edgeC}
\put(90,51){\edgeA \edgeB \edgeC \edgeD \edgeE}
\put(150,51){\edgeC \edgeD \edgeE \edgeF}
\put(60,102){\edgeA \edgeC \edgeD \edgeE \edgeF}
\put(120,102){\edgeE \edgeF \edgeA \edgeB \edgeC}
\put(30,153){\edgeB \edgeC}
\put(90,153){\edgeB \edgeC \edgeD}
\put(150,153){\edgeC \edgeD}
\put(90,85){\edgeB \edgeD \edgeF}

\end{picture}
\caption{A configuration with two fewer than the maximum number of loops.}
\label{oneDefect}
\end{center}
\end{figure}

The smallest change in the number of loops that can be made is to introduce
a defect somewhere in one of the maximal configurations, as shown in
Figure~\ref{oneDefect}.  There are $2N/3$ different such defects that can
be introduced and each reduces the number of loops by 2.

Introducing defects in this way, we can reach all possible configurations.
To see that this is so, we can represent a configuration by labelling the
links on the lattice that do not contain part of a path.  One of every three
links is unoccupied, and every vertex touches one unoccupied link.  These
unoccupied links form a dimer configuration for the vertices of the lattice.
If we draw rhombuses around every dimer and interpret the resulting picture
as the projection of the edges of a stack of cubes, we see that a FPL
configuration is equivalent to a stack of cubes.  Such an identification is
shown in Figure~\ref{cubes}.

\newcommand{\edgeX}{\put(0,0){\line(0,1){34}}}
\newcommand{\edgeY}{\put(0,0){\line(5,-3){30}}}
\newcommand{\edgeZ}{\put(0,0){\line(-5,-3){30}}}

\begin{figure}
\begin{center}
\begin{picture}(300,150)

\thicklines

\put(0,34){
\put(0,0){\edgeB}
\put(30,-17){\edgeB}
\put(60,-34){\edgeA}
\put(0,34){\edgeB \edgeC}
\put(30,17){\edgeA \edgeB \edgeC}
\put(60,0){\edgeA \edgeC}
\put(90,-17){\edgeA}
\put(0,68){\edgeA \edgeB \edgeC}
\put(30,51){\edgeA \edgeC}
\put(60,34){\edgeA \edgeB}
\put(90,17){\edgeA \edgeC}
\put(30,85){\edgeA \edgeB}
\put(60,68){\edgeA \edgeC}
\put(90,51){\edgeB}
\put(60,102){\edgeB}
\put(90,85){\edgeB \edgeC}
\put(120,68){\edgeC}}

\put(150,34){
\put(0,0){\edgeX \edgeY}
\put(30,-17){\edgeX \edgeY}
\put(60,-34){\edgeX}
\put(0,34){\edgeX \edgeY}
\put(30,17){\edgeX \edgeY}
\put(60,0){}
\put(90,-17){\edgeX \edgeZ}
\put(0,68){\edgeY}
\put(30,51){}
\put(60,34){\edgeX \edgeY \edgeZ}
\put(90,17){\edgeZ}
\put(120,0){\edgeX \edgeZ}
\put(30,85){\edgeY \edgeZ}
\put(60,68){\edgeZ}
\put(90,51){\edgeX \edgeY \edgeZ}
\put(120,34){\edgeX \edgeZ}
\put(60,102){\edgeY \edgeZ}
\put(90,85){\edgeY \edgeZ}
\put(120,68){}}

\end{picture}
\end{center}
\caption{An example of the identification of FPL configurations and stacks of
cubes.  One rhombus is drawn centered on each unoccupied link.}
\label{cubes}
\end{figure}
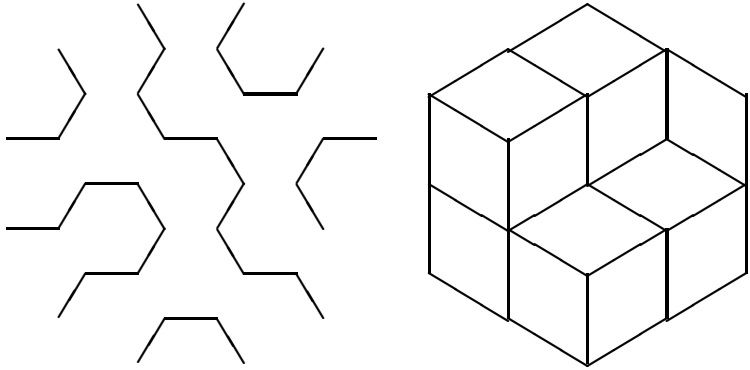

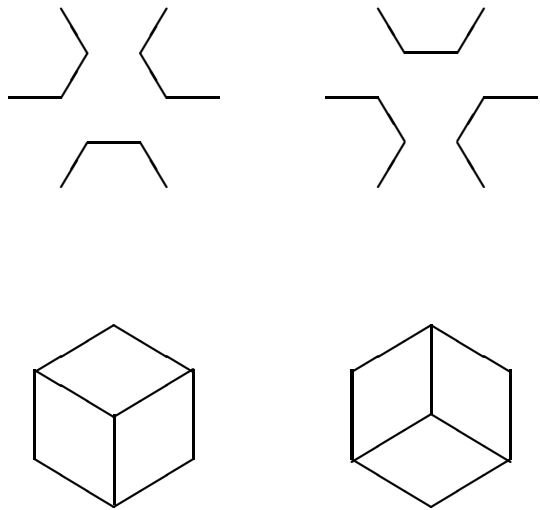
\begin{figure}
\begin{center}
\begin{picture}(200,200)

\thicklines

\put(0,0){\edgeX \edgeY}
\put(30,-17){\edgeX}
\put(0,34){\edgeY}
\put(30,17){}
\put(60,0){\edgeX \edgeZ}
\put(30,51){\edgeY \edgeZ}
\put(60,34){\edgeZ}

\put(120,0){
\put(0,0){\edgeX \edgeY}
\put(30,-17){}
\put(0,34){}
\put(30,17){\edgeX \edgeY \edgeZ}
\put(60,0){\edgeX \edgeZ}
\put(30,51){\edgeY \edgeZ}
\put(60,34){}}

\put(0,120){
\put(0,0){\edgeB \edgeF}
\put(60,0){\edgeD \edgeF}
\put(0,34){\edgeA \edgeC}
\put(60,34){\edgeC \edgeE}
\put(30,17){\edgeA \edgeC \edgeE}}

\put(120,120){
\put(0,0){\edgeB \edgeF}
\put(60,0){\edgeD \edgeF}
\put(0,34){\edgeA \edgeC}
\put(60,34){\edgeC \edgeE}
\put(30,17){\edgeB \edgeD \edgeF}}

\end{picture}
\end{center}
\caption{In the cube representation, introducing a defect is adding or removing
a cube.}
\label{cubeDefect}
\end{figure}

In this new representation, the action of inserting a defect is just the action
of adding or removing a cube.  This identification is exhibited in
Figure~\ref{cubeDefect}.  The result then follows that since every stack of
cubes can be made by adding or removing cubes, every FPL configuration can be
made from one of the maximal ones by inserting some combination of defects.

To obtain an approximation for the free energy, consider first the maximal
state shown in Figure~\ref{noDefects}.  For a lattice of $N$ faces, this
configuration has $(N/3)$ loops. There are 3 such configurations corresponding
to the three-fold translational degeneracy of the state.  To lowest order
then $Z_{FPL} = 3\,n^{N/3}[1+O(n^{-1})]$.  This result was used
in section~\ref{FLsection} to determine the asymptotics of the average loop
length.

Allowing defects, there are $(2N/3)$ locations for a defect and each defect
reduces the number of loops by two.  Defects may be applied in any number and
in any combination, giving the usual sum over disconnected diagrams.  We can
write this as the exponential of the connected diagram (one defect) and we
will be correct except for the effects of excluded volumes which begin with
two-defect connected diagrams and are therefore higher order.  To the next
order, $Z_{FPL} = 3\,n^{N/3} \exp[(2N/3) n^{-2}] \exp[O(n^{-4})]$.

Perturbatively calculating the FPL free energy, we see that
\begin{equation}
F_{FPL}(n) = \frac{1}{3} \log(n) + \frac{2\,n^{-2}}{3} + O(n^{-4}),
\end{equation}
in conformity with Baxter's result shown in equation~(\ref{FFPL}).

This point of view incidentally leads to a simple expression for the
entropy density of the FPL configurations at $n=1$.  At this point, all
configurations are weighted equally and $Z_{FPL}$ is just the number of
configurations, or the exponential of the entropy.  Then calculating the
partition function is just the problem of counting the number of coverings of
the honeycomb lattice by paths, which is the number of different
possible stacks of cubes, which is the old combinatorial problem of counting
plane partitions.  Elser~\cite{Elser} has calculated the asymptotics of plane
partitions for large arrays of numbers.

The result applied to this case is entirely dependent on the shape of the
boundary, even in the thermodynamic limit.  This is to be expected when $n=1$,
because this is in the small-$n$ phase where the model is critical.  For a
lattice of $N$ faces and free boundary conditions, the maximum entropy is
obtained for a hexagon-shaped boundary and in that case the partition function
is asymptotically
\begin{equation}
Z_{FPL}(1) = \exp \left[ N \left( \frac{3}{2} \log 3 - 2 \log 2 \right)
\right].
\end{equation}

\section {Comparison with surface tension}
\setcounter{equation}{0}
\baselineskip 18.5pt

The ground state of the $sl_q(d)$ integrable lattice model is
$(d+1)$-fold degenerate.  This implies the existence
of a notion of interfacial tension  $S(\gamma)$ away from the critical point
between regions of differing antiferromagnetic polarization.  By considering
finite-size corrections, de~Vega~\cite{deVega} has derived transcendental
equations for this interfacial tension and computed the asymptotic behavior
of $S$ in the limits $\gamma \rightarrow 0$ and $\gamma \rightarrow \infty$.

Scaling arguments originally due to Widom~\cite{Widom} predict that the
scaling relation, $S \xi \sim 1$ should hold near the critical point, $\gamma
\rightarrow 0$ or equivalently $n \rightarrow 2^+$.  It would be interesting to
test this relation in this case, but we know of no explicit expression for the
interfacial tension.

Away from the critical point however, a comparison can be made.  The asymptotic
behavior of the interfacial tension for $\gamma \rightarrow \infty$ was
extracted by deVega, and the result is
\begin{equation}
S(\gamma) = \frac{d}{d+1} \gamma + O(1).
\end{equation}
In the case of the FPL model, $d=2$, $n=e^\gamma+e^{-\gamma}$, and
\begin{equation}
S(n) = \frac{2}{3} \log(n) + O(1).
\end{equation}

\begin{figure}
\epsfbox{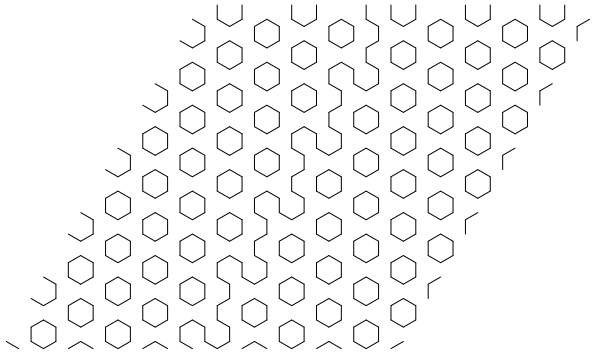}
\caption{An interface separating two regions of differing polarization.}
\label{interface}
\end{figure}

This result may be compared with a perturbative calculation.  Consider the sum
over FPL states at large-$n$ with the constraint that boundary conditions are
fixed to cause frustration in the bulk, as in Figure~\ref{interface}.  The
configuration in that figure has the maximum number of loops possible and
is the analog of the configuration shown in Figure~\ref{noDefects}.  Denoting
the sum over defects in this configuration by $Z'_{FPL}$, the interfacial
tension is defined to be the change in free energy per unit length of the
interface:
\begin{equation}
\frac{Z'_{FPL}}{Z_{FPL}} \sim e^{-LS}
\label{Sdef}
\end{equation}
where $L$ is the vertical size of the lattice.

For a lattice of N faces, the maximum number of loops possible in the presence
of the constraint is $(N/3) - (2L/3)$ instead of $(N/3)$.  The maximal state
in the presence of the constraint is now $3 \times 2^{2L/3}$-fold
degenerate, because there are $(2L/3)$ locations near the interface where
defects may be freely introduced without changing the number of loops.  To
lowest order therefore,
\begin{eqnarray}
Z_{FPL} & \simeq & 3 n^{N/3} \\
Z'_{FPL} & \simeq & 3 2^{2L/3} n^{(N-2L)/3}.
\end{eqnarray}
Reading off the exponents, we have from equation~(\ref{Sdef}) the result that
\begin{equation}
S(\gamma) = \frac{2}{3} \log (n) + O(1).
\label{S}
\end{equation}
Equation~(\ref{S}) is apparently consistent with the large-$\gamma$ asymptotics
derived in~\cite{deVega}.

Equation~(\ref{S}) together with equation~(\ref{xi}) show that $S \xi \neq 1$
in the FPL model.  More generally, from the correlation length calculation it
is clear that for large $\gamma$ the leading behavior of the correlation length
for any value of $d$ will always be $\gamma$, and the leading behavior of $S$
is always  $\gamma d / (d+1)$.

\vskip .5truein
{\Large{\bf Acknowledgement}}

The author is grateful to Professor Nikolai Reshetikhin for many helpful
conversations.

\end{document}